%Paper: hep-th/9507171
%From: ABDALLA ELCIO <elcio@ictp.trieste.it>
%Date: Mon, 31 Jul 1995 09:16:28 +0200 (MET DST)
%Date (revised): Thu, 19 Oct 1995 09:30:13 +0100 (MET)

%%%%%%%%%%%%%%%%%%%%%%%%%%%%%%%%%%%%%%%%%%%%%%%%%%%%%%%%%%%%%%%%%
%\magnification = 1200
%

%
\font\eightrm=cmr8
\font\eighti=cmmi8
\font\eightsy=cmsy8
\font\eightbf=cmbx8
\font\eighttt=cmtt8
\font\eightit=cmti8
\font\eightsl=cmsl8
\font\sixrm=cmr6
\font\sixi=cmmi6
\font\sixsy=cmsy6
\font\sixbf=cmbx6
\catcode`@11
\newskip\ttglue
\font\grrm=cmbx10 scaled 1200

\def\eightpoint{\def\rm{\fam0\eightrm}
\textfont0=\eightrm \scriptfont0=\sixrm \scriptscriptfont0=\fiverm
\textfont1=\eighti \scriptfont1=\sixi \scriptscriptfont1=\fivei
\textfont2=\eightsy \scriptfont2=\sixsy \scriptscriptfont2=\fivesy
\textfont3=\tenex \scriptfont3=\tenex \scriptscriptfont3=\tenex
\textfont\itfam=\eightit \def\it{\fam\itfam\eightit}
\textfont\slfam=\eightsl \def\sl{\fam\slfam\eightsl}
\textfont\ttfam=\eighttt \def\tt{\fam\ttfam\eighttt}
\textfont\bffam=\eightbf
\scriptfont\bffam=\sixbf
\scriptscriptfont\bffam=\fivebf \def\bf{\fam\bffam\eightbf}
\tt \ttglue=.5em plus.25em minus.15em
\normalbaselineskip=6pt
\setbox\strutbox=\hbox{\vrule height7pt width0pt depth2pt}
\let\sc=\sixrm \let\big=\eightbig \normalbaselines\rm}
\newinsert\footins
\def\newfoot#1{\let\@sf\empty
  \ifhmode\edef\@sf{\spacefactor\the\spacefactor}\fi
  #1\@sf\vfootnote{#1}}
\def\vfootnote#1{\insert\footins\bgroup\eightpoint
  \interlinepenalty\interfootnotelinepenalty
  \splittopskip\ht\strutbox % top baseline for broken footnotes
  \splitmaxdepth\dp\strutbox \floatingpenalty\@MM
  \leftskip\z@skip \rightskip\z@skip
  \textindent{#1}\footstrut\futurelet\next\fo@t}
\def\fo@t{\ifcat\bgroup\noexpand\next \let\next\f@@t
  \else\let\next\f@t\fi \next}
\def\f@@t{\bgroup\aftergroup\@foot\let\next}
\def\f@t#1{#1\@foot}
\def\@foot{\strut\egroup}
\def\footstrut{\vbox to\splittopskip{}}
\skip\footins=\bigskipamount % space added when footnote is present
\count\footins=1000 % footnote magnification factor (1 to 1)
\dimen\footins=8in % maximum footnotes per page

\def\ref#1{$^{#1}$}
\def\flex{\raise 6pt\hbox{$\leftrightarrow $}\! \! \! \! \! \! }
\def\oversome#1{ \raise 8pt\hbox{$\scriptscriptstyle #1$}\! \! \! \! \! \! }
\def\tr{ \mathop{\rm tr}}

\newbox\bigstrutbox
\setbox\bigstrutbox=\hbox{\vrule height10pt depth5pt width0pt}
\def\bigstrut{\relax\ifmmode\copy\bigstrutbox\else\unhcopy\bigstrutbox\fi}
\def\refer[#1/#2]{ \item{#1} {{#2}} }
\def\rev<#1/#2/#3/#4>{{\it #1\/} {\bf#2}, {#3}({#4})}
\def\boxit#1{\vbox{\hrule\hbox{\vrule\kern3pt
\vbox{\kern3pt#1\kern3pt}\kern3pt\vrule}\hrule}}

\def\2figure#1#2#3#4{\vbox{ \hrule width#1truecm \hbox{\vrule height#2truecm
\hskip #1truecm
\vrule height#2truecm }\hrule width#1truecm \hbox{\vrule\vbox{\hsize #1truecm
\baselineskip=10pt
\noindent\strut#3}\vrule}\hrule width#1truecm
\hbox{\vrule\vbox{\hsize #1truecm
\baselineskip=10pt
\noindent\strut#4}\vrule}\hrule width#1truecm  }}
\def\3figure#1#2#3#4#5{\vbox{ \hrule width#1truecm \hbox{\vrule height#2truecm
\hskip #1truecm
\vrule height#2truecm }\hrule width#1truecm \hbox{\vrule\vbox{\hsize #1truecm
\baselineskip=10pt
\noindent\strut#3}\vrule}\hrule width#1truecm
 \hbox{\vrule\vbox{\hsize #1truecm
\baselineskip=10pt
\noindent\strut#4}\vrule}
\hrule width#1truecm \hbox{\vrule\vbox{\hsize #1truecm
\baselineskip=10pt
\noindent\strut#5}\vrule}\hrule width#1truecm  }}

\def\sqr#1#2{{\vcenter{\hrule height.#2pt
   \hbox{\vrule width.#2pt height#1pt \kern#1pt
    \vrule width.#2pt}
    \hrule height.#2pt}}}

% Here are my additional definitions:

\def\smin{\,\raise 0.06em \hbox{${\scriptstyle \in}$}\,}
\def\smsubset{\,\raise 0.06em \hbox{${\scriptstyle \subset}$}\,}

\def\Natural{\hbox{\hskip 1.5pt\hbox to 0pt{\hskip -2pt I\hss}N}}

\def\Rational{\hbox{\hbox to 0pt{\hskip 2.7pt \vrule height 6.5pt
                                  depth -0.2pt width 0.8pt \hss}Q}}
\def\Real{\hbox{\hskip 1.5pt\hbox to 0pt{\hskip -2pt I\hss}R}}
\def\Complex{\hbox{\hbox to 0pt{\hskip 2.7pt \vrule height 6.5pt
                                  depth -0.2pt width 0.8pt \hss}C}}
%%%%%%%%%%%%%%%%%
% definitions for the second book
%tcap1

\def \E {{{\rm e}}}

%%%%%%%%tcap2

%%%%%%%%tcap3
\def \1ok{{1\over \kappa ^2} }

\def \3dslim {{\rm DS}\!\!\!\!\!\!\!\!\lim }
\def \4dslim {{\rm DS}\!\!\!\!\!\!\!\!\!\!\lim }
\def \tr {{\rm tr}\, }

\def \2kk{\left( \matrix {2k\cr k\cr }\right) }
\def \Rs4{{R^k\over 4^k} }

%%%%%%%%%tcap4
\def \1ok{{1\over \kappa ^2} }

%%%%%%%%%%tcap5

%%%%%%%%%%tcap6
%\def \DSLim {{\rm DS}\!\!\lim }
%%%%%%%%%%tcap7
%nada
%%%%%%%%%%tcap8
%nada
%%%%%%%%%appb

\magnification 1200
\nopagenumbers

\hfill CERN-TH/95-209

\hfill IC/95/201

\hfill hep-th/9507171
\vskip 1cm

\centerline {\grrm  Gravitational interactions of integrable models}
\vskip .6cm

\centerline {E. Abdalla\newfoot {${}^*$}{Permanent address:
Instituto de F\'\i sica - USP, C.P. 20516, S. Paulo, Brazil.} and M.C.B.
Abdalla\newfoot{${}^\dagger $}{Permanent address: Instituto de
F\'\i sica Te\'orica - UNESP, R. Pamplona 145, 01405-900, S. Paulo, Brazil.}}
\vskip .4cm

\centerline{ CERN-TH}

\centerline{ CH-1211 Geneva 23 - Switzerland}

\vskip .2cm

\centerline { International Centre for Theoretical Physics - ICTP}

\centerline {34100 Trieste, Italy }
\vskip 3cm
\centerline{\bf Abstract}
\vskip .5cm

\noindent We couple non-linear $\sigma$-models to Liouville gravity, showing
that integrability properties of symmetric space models still hold for the
matter sector. Using similar arguments for the fermionic counterpart, namely
Gross--Neveu-type models, we verify that such conclusions must also hold for
them, as recently suggested.
\vfill

\centerline {To appear in Physics Letters B}

\vfill
\noindent CERN-TH/95-209

\noindent IC/95/201

\vskip .3cm
\noindent hep-th/9507171
\vskip .3cm
\noindent July 1995
\eject
\countdef\pageno=0 \pageno=1
\newtoks\footline \footline={\hss\tenrm\folio\hss}
\def\folio{\ifnum\pageno<0 \romannumeral-\pageno \else\number\pageno \fi}
\def\advancepageno{\ifnum\pageno<0 \global\advance\pageno by -1
\else\global\advance\pageno by 1 \fi}

\noindent Non-linear $\sigma$-models defined on a symmetric space $M=G/H$ are
integrable.\ref{1} Moreover they are classically conformally invariant and do
not interact with a gravitational field, which is readily seen to
cancel as we substitute the metric, written in the conformal gauge, into
the Lagrangian.

However, in the quantum theory several new features arise. The first of them
is the mass generation as arising from the constraint due to quantum
fluctuations, which generate a vacuum expectation value for the
Lagrange multiplier. This fact leads to a non-conformally-invariant term,
thus coupling the matter fields explicitly to the
Liouville field. Moreover the trace anomaly in the computation of the
determinant of the d'Alembert operator leads to a Liouville term which has
also analogous contributions from the gravitational ghosts. A further issue
is the fact that, globally, we cannot use the conformal gauge in a general
Riemann surface, leading, in these cases, to extra moduli integrations.
However, we stay, for the moment, in a base space with a trivial global
topology, since the anomaly is generally connected with the Liouville
field.

In order not to overload our formulae we restrain ourselves to the $O(N)$
non-linear $\sigma$-model, or else the $\Complex P^{N-1}$ model. However the
results are trivially generalized to any symmetric space. These examples
are leading, since already in the case of flat space both are
integrable, with the diference that in the former the integrability condition
stays valid in the quantum theory, since the gauge group $H=SO(N-1)$ is simple,
while in the latter a quantum anomaly arises, spoiling integrability.
This fact remains true in a general Grassmannian, where the gauge group
is $H=S(U(N-p)\times U(p))$, or $H=S(O(N-p)\times O(p))$, where  anomalies
generated by the gauge fields $SU(p)\times U(1)$, or $SO(p)$ spoil the
conservation laws. This is an important issue for string theory,
where the relevant quotient space is $SO(32)/SO(8)\times SO(24)$, as we
comment later on.

The partition function for the $O(N)$-model is given by the expression\ref{1}
$$
\eqalignno{
{\cal Z} = & \int {\cal D}\varphi {\cal D}g^{\mu\nu}\E^{i\int {\rm d}^2x
{1\over 2}\sqrt{-g}g^{\mu\nu} \partial _\mu\varphi_i\partial _\nu\varphi_i} \cr
& \times {\cal D}\alpha \E^{i\int {\rm d}^2x \sqrt{-g} {\alpha(x)\over 2\sqrt
N}
\left[\varphi_i^2 - {N\over 2f}\right]} {\cal D}[{\rm ghosts}] \E^{iS_{\rm
grav}[{\rm ghosts}]}\quad . &(1)\cr}
$$

It is defined in terms of a Weyl-invariant action, a non-Weyl-invariant
constraint (due to the field $\alpha (x)$), and a non-Weyl-invariant measure.
The Weyl non-invariance of the measure has been studied in ref. [2]. There,
it has been proved that the scalar fields measure transforms under Weyl
transformation $g'= \E^\sigma g$ as
$$
\prod _{i=1}^N {\cal D}_{\E^\sigma g}\varphi_i = \E^{-{iN\over 48 \pi}S_L}
{\cal D}_{g}\varphi_i\quad ,\eqno(2)
$$
where $S_L$ is the Liouville action.

Since the $\varphi_i$-fields build the $N$-plet appearing asymptotically,
this is the only contribution to the Liouville action beside that of the
ghosts, which gives the usual contribution $-26$. Therefore, writing the metric
as a Liouville factor times a residual metric $\hat g$, we are left with
$$
\eqalignno{
{\cal Z} = & \int {\cal D}\varphi {\cal D}\sigma \E^{i\int {\rm d}^2\xi\left[
{1\over 2} \hat g^{\mu\nu} \partial _\mu\varphi_i\partial
_\nu\varphi_i\right]}
{\cal D}\alpha \E^{i\int {\rm d}^2x \E^{\gamma \sigma} {\alpha(x)\over 2\sqrt
N} \left[\varphi_i^2 - {N\over 2f}\right]}\cr
& \times {\cal D}[{\rm ghosts}^{(0)}] \E^{iS^{(0)}_{{\rm grav}}[{\rm ghosts}]}
\E^{i{26-N\over 24\pi}\int {\rm d}^2\xi \left[{1\over 2} \partial^\alpha
\sigma\partial _\alpha \sigma - \mu \E^{\gamma \sigma} + Q\hat R \sigma\right]}
\quad ,&(3)\cr}
$$
where $\gamma, \mu, Q$ are parameters that include possible quantum
corrections arising from renormalization effects. Actually, we are mostly
interested in the case where the background metric $\hat g^{\mu\nu}$
corresponds to Minkowski space $\eta^{\mu\nu}$. In the $\Complex P^{N-1}$
model we find
$$ \eqalignno{
{\cal Z} = & \int {\cal D}\bar z{\cal D}z{\cal D}\sigma {\cal D}A_\mu{\cal D}
\alpha {\cal D}[{\rm ghosts}^{(0)}]\E^{iS^{(0)}_{\rm grav}[{\rm ghosts}]}\cr
&\times\E^{i\int{\rm d}^2\xi\left[\eta^{\mu\nu}\overline{D^\mu z}D_\nu z+\E^{
\gamma \sigma}{\alpha(x)\over\sqrt N}\left[\bar zz-{N\over 2f}\right]\right]}
\cr
&\times \E^{i{13-N\over 12\pi}\int{\rm d}^2\xi\left[{1\over 2}\partial^\alpha
\sigma\partial_\alpha\sigma -\mu\E^{\gamma\sigma}+Q\hat R\sigma\right]}\quad .
&(4)\cr}
$$

Weak gravitational fields in eq. (3) are formally obtained for $\vert N-26
\vert\to\infty $. The semiclassical gravity is actually obtained for $N-26\to
-\infty$. The constraint displays a gravitational interaction, which is
trivial in the sense that it may be absorbed in the $\alpha$-field measure,
since we may define $\tilde\alpha=\E^{\gamma\sigma}\alpha$, and the Jacobian
$$
J= \det \E^{\gamma\sigma}\eqno(5)
$$
corresponds to a renormalization of the interaction of the Liouville field with
the background curvature, namely $\delta {\cal L} \simeq \hat R \sigma$. Thus
we separate, at the Lagrangian level, three sectors, namely $\sigma$-model,
Liouville and ghost sectors. The $O(N)$ $\sigma$-model sector is integrable,
even before the $\alpha$-field redefinition, because both equations
$$
\eqalignno{
\partial^\mu j_\mu^{ij} & = 0 \quad ,&(6a)\cr
\left[ \partial _\mu + {2f\over N}j_\mu, \partial_\nu +{2f\over N}j_\nu \right]
& = 0 \quad ,&(6b)\cr}
$$
obeyed by the Noether current ${j_\mu}_{ij}=\varphi_i\flex\partial_\mu\varphi_j
$ (or ${j_\mu}_{ij}=z_i\flex D_\mu\bar z_j$ for $\Complex P^{N-1}$), still
hold true classically, independently of the redefinition implied by eq. (5).
However, quantum mechanically, models such as $\Complex P^
{N-1}$, which are defined on a symmetric space $G/H$, where $H$ is not simple,
have an anomaly in eq. (6$b$), which is thus spoiled by quantum effects. For
the $O(N)$-model, $H=O(N-1)$ is simple and eq. (6$b$) holds in the quantum
theory. For the $\Complex P^{N-1}$-model, the gauge group is $H=SU(N-1)\otimes
U(1)$, not simple, and allows anomalous terms. Any symmetric space model $G/H$
with a simple gauge group $H$ displays, in the quantum theory, an infinite
number of conservation laws.\ref{16} If $H$ is not simple the model is
anomalous; only certain couplings with fermions (as e.g. supersymmetric)
render integrability back in the quantum theory.\ref{10} The models
$O(8,8)/O(8)\times O(8)$ and $O(8,24)/O(8)\times O(24)$ are
anomalous.\ref{1,10} Due to cancellation of anomalies in the supersymmetric
case,\ref{1,10} use of the non-local conservation laws may then
be effective.\ref{11,17} The first reanalysis that has to be carried out
in the quantum theory with gravitational fields, is the issue  of the
Wilson expansion of the currents with Liouville fields present; namely, in
order to obtain an infinite number of conservation laws, the first of
them,
$$Q^{(2)}=\int {\rm d}x{\rm d}yJ_0(t,x)\epsilon (x-y) J_0(t,y)- Z\int
{\rm d}xJ_1(t,x)\eqno(7)$$
is well defined and conserved by means of a suitable definition of the
renormalization constant $Z$, which amounts to analysing the
short-distance behaviour of the product of the currents
$$J_\mu(x+\epsilon )J_\nu(x)=C_{\mu\nu}^\rho (\epsilon)J_\rho (x) +
D_{\mu\nu}^{\rho\sigma}(\epsilon)\partial_\rho J_\sigma(x) +
E_{\mu\nu}^{\rho_i}(\epsilon){\cal O}_{\rho_i}(x)\quad ,\eqno(8)$$
verifying that the last term does not really occur, since it spoils the
conservation law.\ref{10} As a matter of fact, the problem is similar to the
one discussed in relation to the pure matter case,\ref{16} as far as one
dresses the fields with the gravitational background, as we discuss below. The
consequence is that if the gauge group is simple, there are contributions
spoiling the higher conservation laws. In the $\Complex P^{N
-1}$ case the anomaly
is given by
$${d Q^{(2)}\over dt}={1\over 2\pi}\int {\rm d}x \E^{\beta\sigma}
z\overline z F_{\mu\nu}\epsilon^{\mu\nu}\quad .\eqno(9)$$
Once more, when the gauge group $H$ is simple, no source of anomaly
arises, since there is no candidate to be dressed. The constant $\beta$ is
fixed imposing that the conformal dimension of the integrand be\ref 6 one).

The theory cannot be completely defined before its constraint structure is
solved. In fact, as in the case of WZW gauge interactions, the constraints
play a crucial role in the definition of asymptotic states; different sectors
are
decoupled at the Lagrangian level, but the first class constraints relate them
by means of the definition of the physical states %%%%%may be the
only coupling to the remaining sectors of the
theory. Since they are first class, they imply a choice of the physical
states of the theory. Such constraints may be obtained by coupling the
theory to external
gravitational fields, as proposed in ref. [3] in the gauged WZW $G/H$-coset
construction. In other words, they coupled to external gauge fields, which
turn out to disappear due to field redefinition, and the variation of the
partition function with respect to such external fields are first-class
constraints! Here we have the analogous construction coupling the matter
fields to a classical gravitational field. For the Liouville action we have
$$
{\cal L}_L= {1\over 4\pi}\sqrt{-g}\left( {1\over 2}g^{\mu\nu} \partial _\mu
\sigma \partial _\nu \sigma+\sigma R(g)-\mu\E^{\sigma}\right)
\quad ,\eqno(10)
$$
which for $g^{\mu\nu}=\E^{\sigma'}\hat g^{\mu\nu}$, and adding the contribution
${\cal L}_L[\sigma',\hat g;\mu =0]$ arising from the matter/ghost system
with a definite choice of renormalization, leads to
$$
\eqalignno{
{\cal L}_L^{{\rm tot}}& = {\cal L}_L[\sigma,\E^{\sigma'}\hat g^{\mu\nu}] +
{\cal L}_L[\sigma',\hat g^{\mu\nu}] \cr
& = {1\over 4\pi}\sqrt{-\hat g}\left( {1\over 2}\hat g^{\mu\nu} \partial _\mu
(\sigma + \sigma')\partial _\nu (\sigma+ \sigma') + (\sigma + \sigma')
R(\hat g) - \mu \E^{\sigma + \sigma'}\right) \cr
& \equiv  {\cal L}_L [\sigma + \sigma', \hat g]\quad ,&(11)\cr}
$$
which is the analogue of the Polyakov--Wiegmann identity.\ref{4} Therefore the
partition function does not depend on $\sigma'$, which leads to a
(first-class) constraint, corresponding to the Wheeler--de Wit equation.

First-class constraints are realized as equations defining the physical states.
The dynamics, on the other hand, is obtained from the corresponding
(factorized) Lagrangian, which is equivalent to eq. (6$b$), implying, in turn,
higher conservation laws, and a factorizable $S$-matrix. Moreover, they imply
also a Yangian-type algebra as described in ref. [5], and consequently a
half-affine algebra, by means of a Lie--Poisson action.

The Wheeler--de Wit equation corresponds to the vanishing Hamiltonian of the
composite matter--Liouville--ghost system when acting in a physical state. If
we work on a fixed background where the matter fields are constrained $(\vec
\varphi^2=1)$ we obtain, for the Wheeler--de Wit equation (see p. 44 of ref.
[6]):
$$
\left[\left({\partial\over\partial\sigma}+{Q\over 2}\right)^2+\sum_{i=1}^N{
\partial^2\over\partial\varphi_i^2}+2-\left({Q\over 2}\right)^2\right]\tilde
\psi=0\quad.\eqno(12)
$$

For $\tilde \psi= \E^{-{Q\over 2}\sigma}\psi$, we obtain the
$(N+1)$-dimensional Helmholtz equation on a cylinder of unit radius (we
suppose ${N\over 2f}=1$), that is
$$
\left[ {\partial^2 \over \partial \sigma^2} + \sum_{i=1}^N {\partial^2\over
\partial \varphi_i^2 }+2-\left({Q\over 2}\right)^2\right]\psi=0\quad.\eqno(13)
$$

With the ansatz
$$
\psi = \E^{-\beta\sigma}\chi(\varphi_i)\quad ,\eqno(14)
$$
we obtain
$$
\left[\sum_{i=1}^N \partial _i^2 + 2-\left({Q\over 2}\right)^2 + \beta^2\right]
\chi\equiv \left[\sum_{i=1}^N \partial _i^2 + m^2\right]\chi = 0\quad,
\eqno(15) $$
on an $(N-1)$-dimensional sphere. This equation is solvable in terms of the
Gegenbauer polynomials for\ref{7}
$$
m^2 = l(N+l-2)\quad ,\eqno(16)
$$
where $l$ is an integer. For $l=0$ the solution is a constant and corresponds
to the matter vacuum. Using
$$
Q^2 = {25-N\over 3}\quad ,\eqno(17)
$$
we obtain
$$
\beta = \sqrt{{19\over 3} + (l-{1\over 3})N + l(l-2)}\quad,\eqno(18)
$$
which is real for $l \ge 1$. Although simple, the dressing involves also an
oscillatory term for $N\ge 20$, if $l=0$, or $N\ge 26$.

The case of fermionic interactions can be dealt with similarly. The analysis
of the chiral Gross--Neveu model was performed in [10]  (see also [18]).
There is no candidate to the anomaly term either, obeying the usual symmetry
requirements. The only difference is that
in the case of the Gross--Neveu model, the bare fermion has to be further
dressed by an explicit Liouville exponential, since it has a non-trivial
conformal dimension, that is, we need to define $\psi$ as
$$
\psi_{{\rm bare}} = \E^{{1\over 4}\sigma}\psi\quad ,\eqno(19)
$$
after which we also obtain, for the redefined matter Lagrangian:
$$
{\cal L}= \bar \psi i\not \! \partial \psi + {1\over 2g}\phi^2 - \phi \bar
\psi\psi\quad,\eqno(20)
$$
where $\phi$ was also dressed as $\phi_{{\rm bare}}=\E^{{1\over 2}\sigma}\phi$.
This latter dressing is similar to the dressing of the Lagrange multiplier
in the $\sigma$-model case.
The equations for the Noether current, $j^{ij}_\mu=\bar \psi^j \gamma_ \mu
\psi^i$, namely conservation and pseudo-current divergence equations, lead to
the integrability condition and higher conservation laws since the only
influence of the $\phi$-field is through its relation with the $\psi$ field,
obtained from the Ward identities, corresponding to the $\phi$ equations of
motion. Thus, quantum integrability holds true in this case as well, confirming
ref. [9], where
the integrability of the Gross--Neveu model coupled to gravity has been
recently conjectured, using completely different methods.

A final remark concerns the issue of the infinite Yangian symmetry of the
theory.
The higher symmetries are described in terms of the non-local conserved
charges  of the theory $Q^{(n)}$, studied by several authors.\ref{1,5,8,12}
When appropriately defined (adding combinations of charges of lower genus)
we find an algebra of the type
$$
\{ Q_a^{(m)}, Q_b^{(n)}\} =\tr\tau^a\tau^b Q^{(n+m)}-\sum_{i=0}^{m-1}\sum_{j=0}
^{n-1}\tr\left(\tau^a Q^{(i)}Q^{(j)}\tau^bQ^{(m+n-i-j-2)}\right)\quad.\eqno(21)
$$

The above algebra is of the Yangian type,\ref{13} as shown in refs. [5,14].
Yangians correspond to quantum group symmetries of many integrable
models.\ref{15}
%%%%%%%%%%%%%%%%%%%%%%%%%%%%%%%%%%%%%%%%%%%%%%
The Poisson-algebraic structure dictated by the quantum non-local charges in
the $O(N)$-invariant case is valid for the matter sector, implying a
``half"-affine algebra structure, started out of the $O(N)$ generator
$Q^{(0)}_{ij}$,
which acts locally, that is on a given field $A$:
$$
\delta^{(0)}_{ij} A = \{Q^{(0)}_{ij}, A\}\quad ,\eqno(22)
$$
while for the first non-trivial higher action we follow ref. [8] and define
$$
\delta^{(1)}_{ij} A = \{Q^{(1)}_{ij},A\} + c\left(Q^{(0)}_{ia}\delta^{(0)}_{aj}
- Q_{ja}^{(0)} \delta^{(a)}_{ai}\right)A\quad ,\eqno(23)
$$
where $c$ is a constant to be adjusted\ref{5,8} and one obtains an algebra
which, as claimed, is half of the affine structure. Due to the above algebraic
structure, the issue of conservation of the first charges flows to the higher
ones, rendering the previous discussion quite general.
%%%%%%%%################
However, the symmetric transformations as generated by the
non-local charges are obtained from a Lie--Poisson action,\ref{15} and not by
the familiar Hamiltonian action. For the first few charges, this is exemplified
in eq. (23), and the symmetry transformation satisfies
$$
[\delta^{(m)}_{ij}, \delta^{(n)}_{kl}] = \left( \delta \circ \delta^{(m+n)}
\right)_{ij,kl} \quad ,\eqno(24)
$$
with the obvious notations for the $O(N)$ Lie-algebra indices $i,j,k,l$.
However, we cannot find a Hamiltonian generator $G^{(n)}$ that realizes the
symmetry action $\delta^{(n)}\phi = \{ H^{(n)}, \phi\}$.

Generalization for models on arbitrary symmetric spaces is immediate. If the
gauge group $H$ is simple there is no quantum anomaly and one can essentially
proceed as in the $O(N)$ case. This happens to be analogous in the case of
super $\Complex P^{N-1}$, where the anomaly cancels.\ref{10} More complicated,
purely bosonic models are, generally speaking, anomalous, as e.g. Grassmannian
non-linear $\sigma$-models, where $G=SU(N)$ and $H=S(U(p)\otimes U(N-p))$.

Recently, there have been proposals\ref{11} to explain the string theory
content using the algebraic structure contained in the higher conservation
laws and the previously mentioned half-affine Lie algebra structure. In view
of the above results, the discussion should be pursued, in the quantum theory,
only in cases where there is no anomaly. Therefore we are forced to
go to the supersymmetric case, which is in fact the most interesting as well,
and where the anomaly generally cancels.\ref{10}

Finally, we have to mention that in the quantum theory there are also
field configurations not obeying the constraint ${\vec \varphi}^2=1$.
Thus quantum states smear off such a background, leading to a perturbation
of the form $\delta H= {1\over 2\sqrt N}\tilde\alpha(x)({\vec\varphi}^2-1)$.
For asymptotic fields arising from the integrable $O(N)$ model, it is
reasonable to assume that this corresponds to a scalar field mass term,
namely we substitute the field $\tilde \alpha(x)$ for its vacuum
expectation value, which renders the resulting Wheeler--de Wit equation
still separable and solvable.

\vskip 1cm
\noindent {\bf Acknowledgements}

\noindent The work of M.C.B.A. was partially supported by the World Laboratory.
\vskip 1cm
\penalty-300
\centerline {\bf References}
\vskip 1cm
\nobreak

\refer[[1]/E. Abdalla, M.C.B. Abdalla and K. Rothe, {\it Non-perturbative
methods in two-\-dimen\-sional quantum field theory} (World Scientific,
Singapore, 1991)]

\refer[[2]/J. Distler and H. Kawai, Nucl. Phys. {\bf B 321} (1988) 171]

\refer[/F. David, Mod. Phys. Lett. {\bf A 3} (1988) 1651]

\refer[[3]/D. Karabali and H.J. Schnitzer, Nucl. Phys. {\bf B 329} (1990) 649]

\refer[[4]/A. Polyakov and P. Wiegmann, Phys. Lett. {\bf B 131} (1983) 121;
{\bf B 141} (1984) 223]

\refer[[5]/E. Abdalla, M.C.B. Abdalla, J.C. Brunelli and A. Zadra, Commun.
Math. Phys. {\bf 166} (1994) 379]

\refer[[6]/E. Abdalla, M.C.B. Abdalla, D. Dalmazi and A. Zadra, {\it Lecture
Notes in Physics}, Vol. m20 (Springer-Verlag, Heidelberg, 1994)]

\refer[[7]/N. Ya. Vilenkin, {\it Fonctions sp\'eciales et th\'eorie de la
repr\'esentation des groupes} (Dunod, Paris, 1969)]

\refer[[8]/A. Le Clair and F.A. Smirnov, Int. J. Mod. Phys. {\bf A 7} (1992)
2997]

\refer[[9]/A. Bilal and I. I. Kogan, Nucl. Phys. {\bf B 449} (1995) 569]

\refer[[10]/E. Abdalla, M.C.B. Abdalla and M. Gomes, Phys. Rev. {\bf D 27}
(1983) 825; E. Abdalla, M. Forger and A. Lima-Santos, Nucl. Phys {\bf B256}
(1985) 145.]

\refer[[11]/J.H. Schwarz, Nucl. Phys. {\bf B 447} (1995) 137; {\it Classical
duality symmetries in two dimensions}, CALT-68-1994, hep-th/9505170;]

\refer[/A. Sen, Nucl. Phys. {\bf B 447} (1995) 62]

\refer[[12]/K. Polmeyer, Commun. Math. Phys. {\bf 46} (1976) 207]

\refer[/M. L\"uscher and K. Pohlmeyer, Nucl. Phys. {\bf B 137} (1978) 46]

\refer[/E. Brezin, C. Itzykson, J. Zinn Justin and J.B. Zuber, Phys. Lett.
{\bf B 82} (1979) 442]

\refer[[13]/V.G. Drinfeld, Sov. Math. Dokl. {\bf 32} (1985) 254]

\refer[/O. Babelon and D. Bernard, Commun. Math. Phys. {\bf 149} (1992)
279]

\refer[[14]/D. Bernard, Commun. Math. Phys. {\bf 137} (1991) 191]

\refer[/A. Le Clair and F.A. Smirnov, Int. J. Mod. Phys. {\bf A 7} (1992) 2997]

\refer[/H.J. de Vega, H. Eichenherr and J. M. Maillet,  Nucl. Phys. {\bf
B 240} (1984) 377, Phys. Lett. {\bf B 132 } (1983) 337, Commun. Math. Phys.
{\bf 92} (1984) 507.]

\refer[[15]/D. Bernard, NATO ASI series B, Physics, Vol. 310, ed. L. Bonora et
al. (Plenum Press, New York, 1993)]

\refer[[16]/E. Abdalla, M. Forger and M. Gomes, Nucl. Phys. {\bf B210} [FS6]
(1982)
181]

\refer[[17]/ J. Maharana, Phys. Rev. Lett. {\bf 75} (1995) 205]

\refer[[18]/E. Abdalla and A. Lima-Santos,  Rev. Bras. Fis. {\bf 12} (1982)
293.]

\end